\documentclass[twocolumn,showpacs,preprintnumbers,amsmath,amssymb,pre]{revtex4}

\usepackage{graphicx,color}
\usepackage{dcolumn}
\usepackage{bm}

\begin{document}

\title{Fluctuations in Chemical Gelation }

\author{Kenji Ohira, Masatoshi Sato and Mahito Kohmoto}
\affiliation{
The Institute for Solid State Physics, the University of Tokyo \\
5-1-5 Kashiwanoha, Kashiwa, Chiba 277-8581, Japan
}

\date{\today}

\begin{abstract}
We study a chemical gelation model in two dimensions which includes both
 monomer aggregations and bond fluctuations. Our numerical simulation
 shows that a sol-gel transition occurs when an initial monomer
 concentration is above a critical
 concentration. Fractal aggregates grow until the sol-gel transition
 occurs. After the gelation, however, bond fluctuations break the
 fractal structure and a novel inhomogeneous gel fibre network appears
 instead. A pore size distribution of
 the inhomogeneous structure shows the existence of hierarchical
 structures in the gel phase. It is also found that slow 
 dynamics appear near the critical concentration. 
\end{abstract}

\maketitle


\section{Introduction}
One of the most important characteristics of chemical gels is formation
of heterogeneous structure. 
The structural inhomogeneities
affect their physical properties such as permeability,
elasticity, and optical properties \cite{Gelshandbook}.
Although many studies have been made \cite{Bastide,History}, its
dynamical origin has not been elucidated.
The difficulty is that they are determined by non-equilibrium
dynamics.
Since chemical gels are formed by irreversible aggregations between
constituent monomers, the structures are frozen in the non-equilibrium 
gelation processes.

On the other hand, equilibrium properties of gels have been
investigated with percolation models \cite{Stauffer} which generate equilibrium
ensembles. For example, it is well-known that physical gels formed under annealed
conditions are described by equilibrium
systems such as correlated percolation models \cite{deGennesbook,Coniglio,Gujrati}. However, it
is not obvious that equilibrium systems can describe the quenched
randomness produced by the irreversible aggregations of chemical gels \cite{Martin2}.

\begin{figure}[b]
\includegraphics[width=3cm]{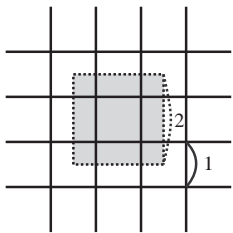}
\caption{A  monomer on the square lattice.}
\label{fig:monomer}
\end{figure}

In this paper, we give a chemical gelation model which
includes both monomer irreversible aggregations and bond
fluctuations, and numerically study its gelation dynamics in two
dimensions.
The model introduces bond fluctuations in a similar manner as the
bond fluctuation model (BFM) \cite{BFM} which has been used to model the Rouse 
dynamics of polymers, and it can simulate polymers and cross-linked gels
in an unified framework. 
By the numerical simulations, a dynamical origin of the inhomogeneities
of chemical gels is reported.
We shall show that bond fluctuations drastically change the structures of gels.
Although the fractal nature is important in growth
kinetics, bond fluctuations break it and lead to a novel inhomogeneous
gel fibre network structure.

The paper is organised as follows. 
Section \ref{sec:model} describes the model and the procedure of
numerical simulations.
In Sec.\ref{sec:results} we present our numerical results and discuss
the gelation dynamics. 
We show that a critical concentration of gelation exists 
due to competition between fractal aggregations and bond fluctuations.
The inhomogeneous gel networks are investigated in detail and
hierarchical structures of the networks are found.
We also report that the cluster size distributions below the critical
concentration broaden and have a tail with large clusters near the
critical concentration.
Finally, concluding remarks are given
in Sec.\ref{sec:conclusion}.

\section{The Model}\label{sec:model}

\begin{figure}[b]
\includegraphics[width=8cm]{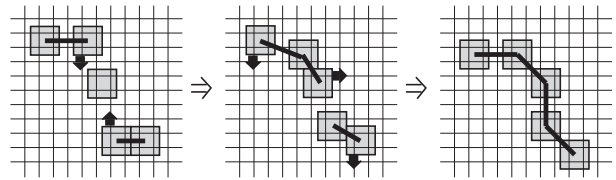}
\caption{An aggregation process of two clusters. Gray plaquettes
 represent monomers and thick lines represent bonds between
 them. Arrows show randomly chosen directions in which monomers move
 within $\Delta t =1$. In the final figure, the bond formation condition
 is met. The two clusters are joined together and become one cluster.}
\label{fig:dynamics}
\end{figure} 

\begin{figure*}[!t]
\includegraphics[width=15cm]{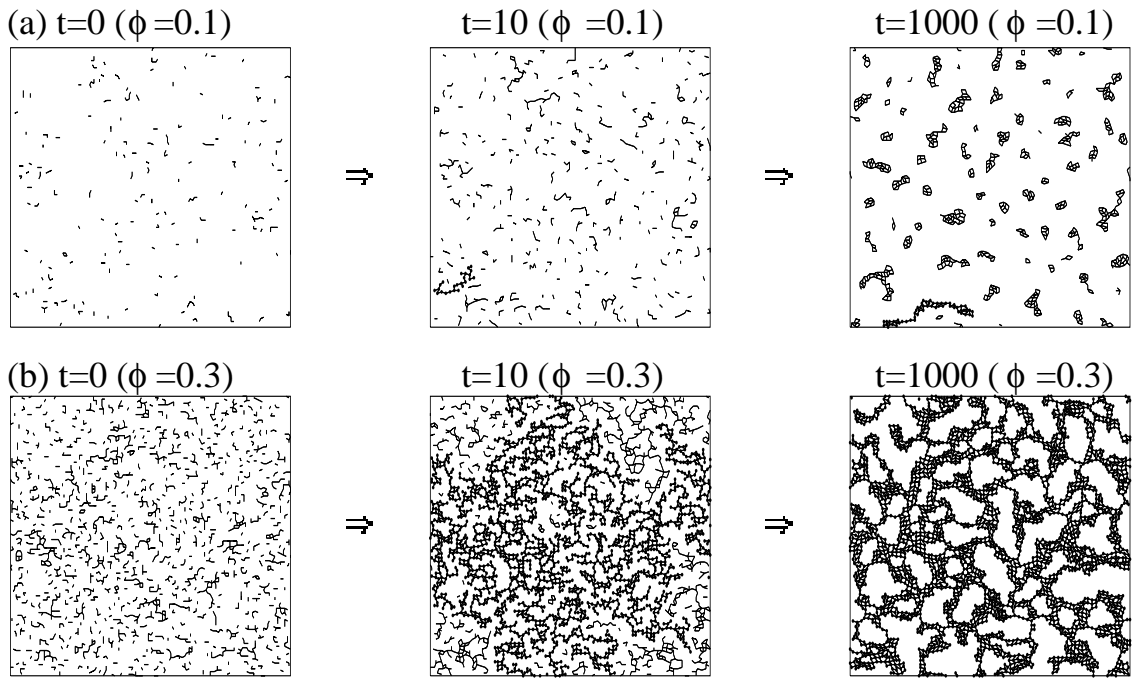}
\caption{Snapshots of gelation processes for (a) $\phi=0.1$ and (b)
 $\phi=0.3$. They are typical examples for high concentrations and low
 concentrations, respectively.  
Thin lines represent bonds and black dots represent monomers in
 the largest cluster. 
For low concentrations ((a)), clusters have globular
 structures, and no sol-gel transition occurs even after $t=1000$.  
For high concentrations ((b)), a fractal cluster forms a
 gel ($t=10$).
After the gelation, however, the intra-cluster bond formation proceeds and
the gel becomes inhomogeneous ($t=1000$).
}
\label{fig:snapshot}
\end{figure*} 

Let us consider $N$ monomers on the $L\times L$ square lattice with the
periodic boundary condition. 
Each monomer is represented by a  plaquette shown in Fig.\ref{fig:monomer} 
and it can jump by one unit lattice. The monomer concentration is given by
\begin{eqnarray}
\phi=\frac{4N}{L^2}. 
\end{eqnarray}

The elementary process of aggregation in our model is formation of a
bond between monomers. 
The bond is created when distance
$l_a$ between two monomers satisfies
\begin{eqnarray}
2\ \le  l_a \le 3,\ &  &  ({\rm namely }\ l_a = 2,\ \sqrt{5},\ 3) 
\label{eqn:la}
\end{eqnarray}
and they both have less than $f$ bonds. ($f$ is the functionality of the
monomer.)
The lower bound of $l_a$ is determined so as to avoid an overlap between
monomers, and the upper bound is the maximal value which guarantees that
the bonds never cut through each other during the course of the
simulation.
As an additional restriction, we forbid formation of a triangle
bond where three monomers bond with each other since it leads to an
artificial triangle-based ladder structure which prohibits isotropic
growth. A bond length $l_b$ can change as long as it satisfies 
\begin{eqnarray}
2\le l_b \le \sqrt{10}, \label{eq:bond_length}
\end{eqnarray}
but a bond itself cannot be broken once it is created.
The value $l_b$ is determined in a similar manner as $l_a$.
The variable bond lengths allow movements of clusters. 
We illustrate an aggregation process of two clusters in
Fig.\ref{fig:dynamics}.

We follow the iterative procedure:
At $t=0$, 
we randomly put $N$ monomers on the lattice avoiding overlaps.
Then for any pair of monomers we create a bond if the distance between
them is one of $l_a$'s in Eq.(\ref{eqn:la}).
Next we randomly choose a monomer and move it to a
randomly chosen direction by one unit lattice
if the monomer does not overlap with others and if the
movement is compatible with the bond length restriction
(\ref{eq:bond_length}). If these conditions are not satisfied, we
proceed with the next iteration without a movement. 
After the movement, we create new bonds if the conditions above are
met, and proceed with the next iteration.
$N$ iterations correspond to a unit physical
time $\Delta t=1$.
As iterations go, monomers form clusters, and  
when one of the clusters is bound to itself via the periodic
boundary condition, we regard that a gel forms and a sol-gel transition occurs.

The algorithm here is a generalization of BFM so
as to include aggregation processes. Indeed, our model reduces to BFM
if we take $f=2$, although the bond
fluctuation region $l_b$ is smaller than that of the original BFM
because creation of bonds with length $l_a$ bring an additional means of
bond crossing. In this paper, we take $f=4$ corresponding to the
functionality of a typical
crosslinker of cross-linked polymer gels, $N,N$-methylenebis. As well as
BFM, each bond in our model effectively represents a group of
consecutive C-C bonds along the backbone of the chemically realistic chain.
In the same way, each monomer is a coarse-grained
monomer which consists of a crosslinker and polymers attached to it.

Simulations are performed for concentrations $\phi=0.1-0.3$ with the
system size $L=200$ unless stated otherwise. 

\section{Results}\label{sec:results}

\subsection{existence of a critical concentration}\label{subsec:exist}

Our simulations show a critical concentration $\phi_{\rm g}$
below which no sol-gel transition occurs.
Typical snapshots are shown in Fig.\ref{fig:snapshot}.
In Fig.\ref{fig:snapshot}(a) the monomer concentration is
$\phi=0.1(<\phi_{\rm g})$ and in Fig.\ref{fig:snapshot}(b) 
$\phi=0.3(>\phi_{\rm g})$.
For $\phi<\phi_{\rm g}$, the
system does not show a sol-gel transition even after very long
iterations ($t\le 10000$), and we obtain a large number of small
globular clusters after all.
On the other hand, the system with $\phi>\phi_{\rm g}$ shows a
sol-gel transition (Fig.\ref{fig:snapshot}(b)). 
The largest cluster in the system becomes self-connected via
the periodic boundary condition and a gel forms in this case.
\begin{figure}[t]
\begin{center}
\includegraphics[width=7cm]{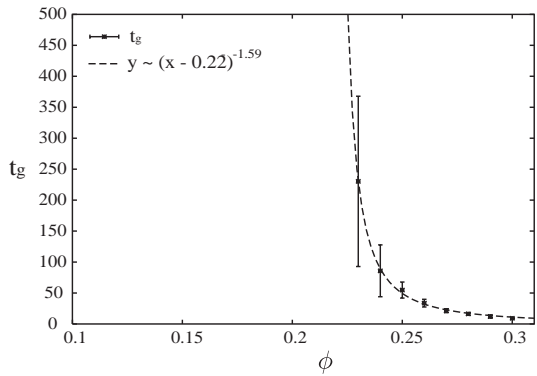}
\caption{
Gelation time $t_{\rm g}$ for various concentrations.
The dashed line represents the best fitted curve for data.
The gelation time shows a divergence at the critical
 concentration $\phi_{\textrm{g}}\simeq 0.22$.
}
\label{fig:tg_c}
\end{center}
\end{figure}
To determine $\phi_{\rm g}$ we plot the gelation time $t_{\rm g}$ as a
function of $\phi$ in Fig.\ref{fig:tg_c}. 
We find that $t_{\rm g}$ increases as $\phi$ decreases and it becomes
infinity at $\phi_{\rm g}\simeq 0.22$ as
\begin{eqnarray}
t_{\rm g}\sim \left(\phi-\phi_{\rm g}\right)^{-\omega},
\end{eqnarray} 
where the power index is $\omega \simeq 1.6$. 
Below $\phi_{\rm g}$, no sol-gel transition takes place. 
The existence of the critical concentration is consistent with experimental
results for cross-linked polymer gels.

\begin{figure}[t]
\begin{center}
\includegraphics[width=7cm]{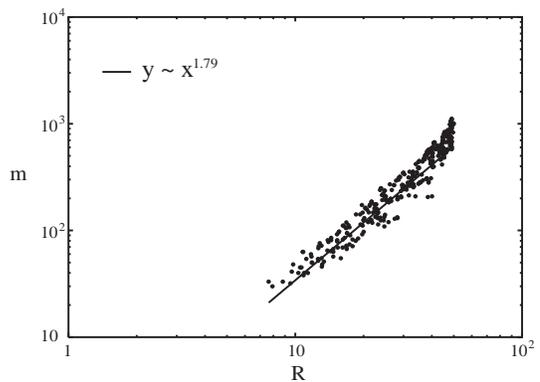}
\caption{
Radius $R$ of the largest clusters versus their mass $m$ at various times
before the sol-gel transition. Here $\phi=0.25 (>\phi_{\rm g})$.
The fractal dimension of the clusters is $d_{\rm f}=1.79\pm 0.06$.}
\label{fig:rg_s_tg}
\end{center}
\end{figure}

For $\phi>\phi_{\rm g}$, the fractal dimension $d_{\rm f}$ of clusters
before gelation is less than
the dimension of the system $d$. 
To show this, we plot in
Fig.\ref{fig:rg_s_tg} the radii $R$ of the largest clusters in the
system at various $t$'s before gelation as a function of their mass $m$.
Here $m$ is defined by the number of monomers in the cluster and $R$ is
defined by 
\begin{eqnarray}
R^2=\frac{1}{2m^2}\ \sum_{i,j}^{m}\ ({\bm r}_i-{\bm r}_j)^2,
\end{eqnarray}
where ${\bm r}_i$ is the position vector of the $i$-th monomer in the cluster.
For the initial monomer concentration $\phi=0.25 (>\phi_{\rm g})$,
we find a power law behaviour 
\begin{eqnarray}
m\sim R^{d_{\rm f}},
\end{eqnarray}
where $d_{\rm f}=1.79\pm 0.06$. 
The fractal dimension $d_{\rm f}$
is almost the same as the corresponding result of the
diffusion limited cluster-cluster aggregation (DLCA) model
in two dimensions, $d_{\rm f}^{\rm DLCA}=1.75\pm0.07$ at the same
concentration \cite{Meakin,Kolb,KH}.
On the other hand, for $\phi< \phi_{\rm g}$, the fractal dimension $d_{\rm f}$ 
is almost the same as the dimension of the system. 
In this case, formation of intra-cluster bonds, namely bond formation
between monomers in the same cluster,  
proceeds before gelation and it makes clusters globular as is seen
in Fig.\ref{fig:snapshot}(a) at $t=1000$.  

The difference of the fractal dimensions naturally explain the
existence of the critical concentrations $\phi_{\rm g}$ as follows. 
When the average mass and radius of clusters are $\langle m \rangle$
and $\langle R \rangle$, respectively, the volume occupied by the
clusters $V_{\rm o}$ is given by
\begin{eqnarray}
V_{\rm o}\sim \left(\frac{N}{\langle m \rangle}\right) {\langle R \rangle}^d
\sim N \langle R \rangle ^{d-d_{\rm f}}.
\label{eq:overlap}
\end{eqnarray}
(Here we have used the relation 
$\langle m \rangle \sim \langle R \rangle ^{d_{\rm f}}$.)
If $V_{\rm o}$ becomes comparable with the volume of the system
\begin{eqnarray}
V_{\rm o} \sim L^{d},
\label{eq:overlapping}
\end{eqnarray}
the clusters begin to overlap and a so-gel transition occurs
by a process similar to the percolation \cite{Gimel}. 
For $\phi>\phi_{\rm g}$, the condition of overlapping (\ref{eq:overlapping}) is
always fulfilled when $\langle R \rangle$
becomes large enough because the inequality $d_{\rm f}<d$ holds.
On the contrary, for $\phi<\phi_{\rm g}$, the condition
(\ref{eq:overlapping}) is not satisfied.
Because of the equation $d \sim d_{\rm f}$, 
$V_{\rm o}$ rarely depends on $\langle R \rangle$ and never exceeds 
the volume of the system. Thus no sol-gel transition occurs in this case.

\begin{figure}[!t]
\begin{center}
\includegraphics[width=8cm]{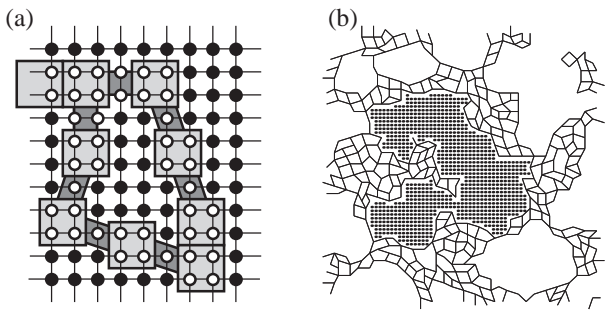}
\caption{(a)A schematic representation of a pore. Light and deep gray
 plaquettes represent monomers and bonds, respectively. Black dots are
 put on the sites uncovered by the monomers and the bonds. We define a
 pore as a cluster which consists of the black
 dots. Figure (b) shows one of the pore. Thin lines represent bonds
 and the dots are the same as that described above. The size of the pore
 is defined by the number of the dots}
\label{fig:pore_exam}
\end{center}
\end{figure}

\subsection{an inhomogeneous structure}\label{subsec:inhomo}
The most remarkable feature of our model is the final structure of gels.
See Fig.\ref{fig:snapshot}(b). As shown in Fig.\ref{fig:rg_s_tg}, gels
have a fractal structure similar to the DLCA model until gel formation.
Indeed, a snapshot at $t=10$ in Fig.\ref{fig:snapshot}(b) also shows
the fractal structure.
However, the fractal structure does not remain after the gel formation.
Structure of gels drastically changes at $t=1000$ in
Fig.\ref{fig:snapshot}(b). 
This is due to formation of intra-cluster bonds after the gel formation.
The formation of intra-cluster bonds breaks the fractal
structure and makes a novel inhomogeneous gel fibre network structure
instead. 
\begin{figure}[!t]
\includegraphics[width=7cm]{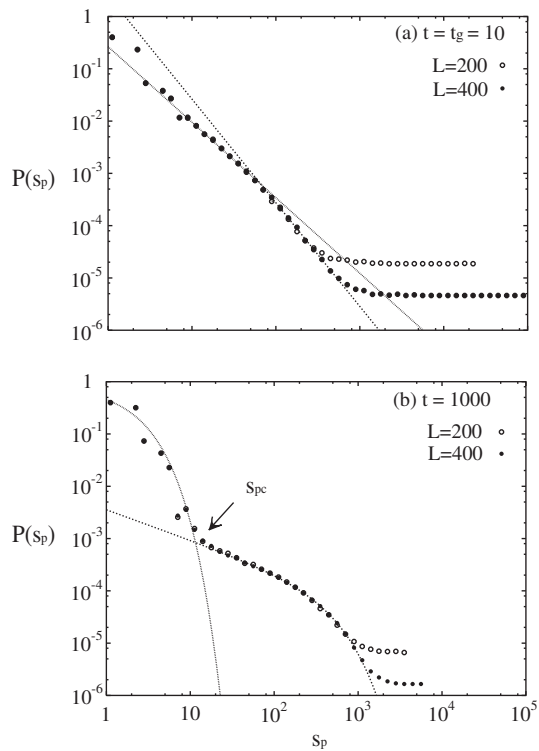}
\caption{Pore size distributions $P(s_{\rm p})$ for $\phi =0.3$. (a) At
 the gelation time, the distribution shows power law decay 
$P(s_{\rm p}) \sim s_{\rm p}^{-\tau}$ with
 $\tau = 1.444 \pm 0.027$ for small $s_{\rm p}$ 
and with $\tau = 1.982 \pm 0.030$ for
 large $s_{\rm p}$. (b) At $t =  1000$, however, the distribution
 decays exponentially. The crossover size $s_{\rm p c}$ separates the
 distribution into monomer-rich and monomer-poor regions.}
\label{fig:pore}
\end{figure}
\begin{figure}[h]
\includegraphics[width=4cm]{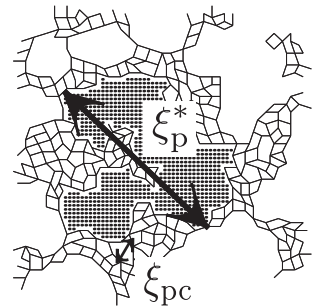}
\caption{Two characteristic size scales $\xi_{\rm pc}$ and 
$\xi_{\rm p}^{\ast}$}
\label{fig:pore_xi}
\end{figure}
\begin{figure}[h]
\includegraphics[width=7.5cm]{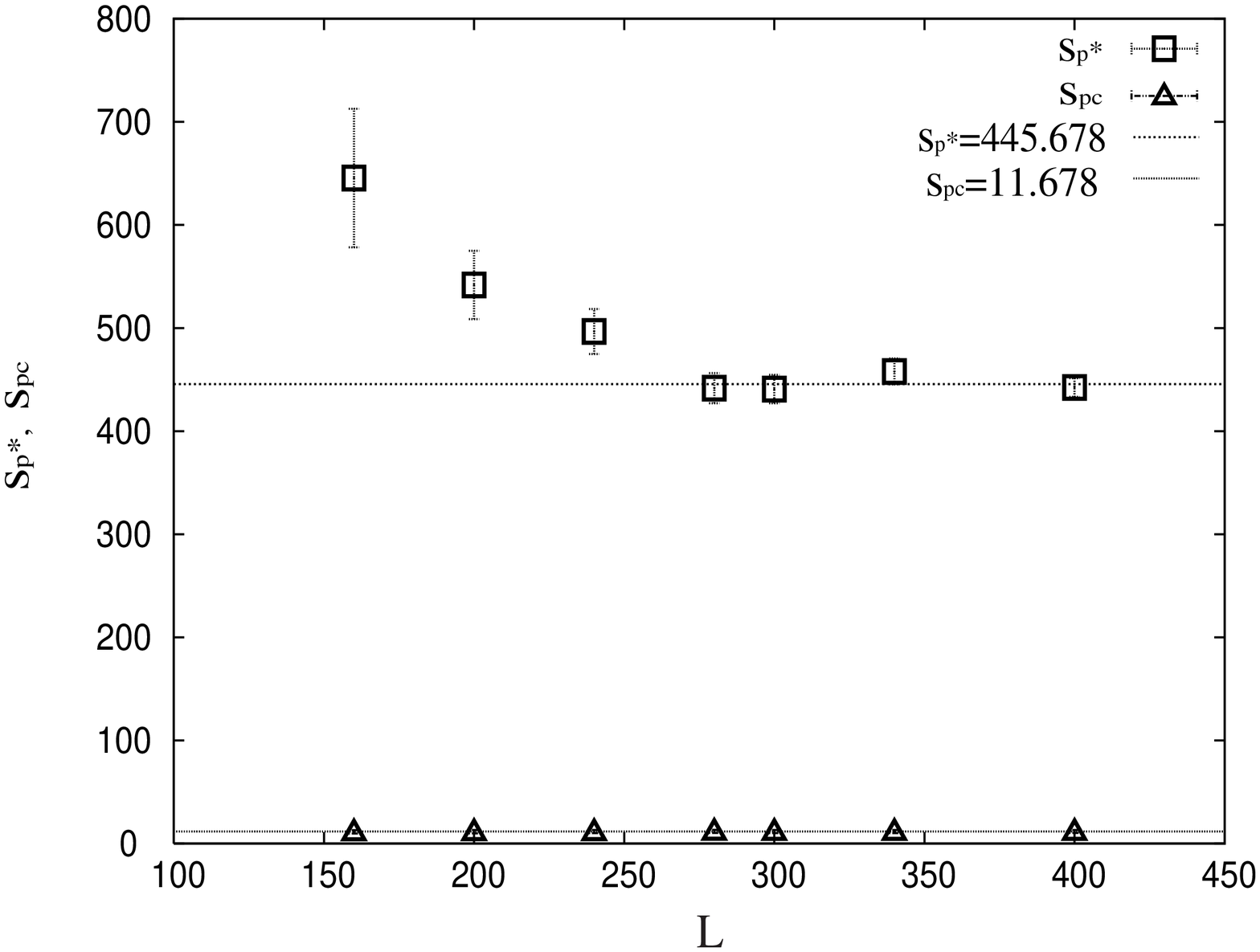}
\caption{Size dependence of two characteristic scales $s_{\rm pc}$ and 
$s_{\rm p}^{\ast}$ of inhomogeneity for $\phi=0.3$ at 
$t=1000$. They rarely depend on $L$ for large $L$.}
\label{fig:pore_size_eff}
\end{figure}
\begin{figure*}[hbt]
\includegraphics[width=15cm]{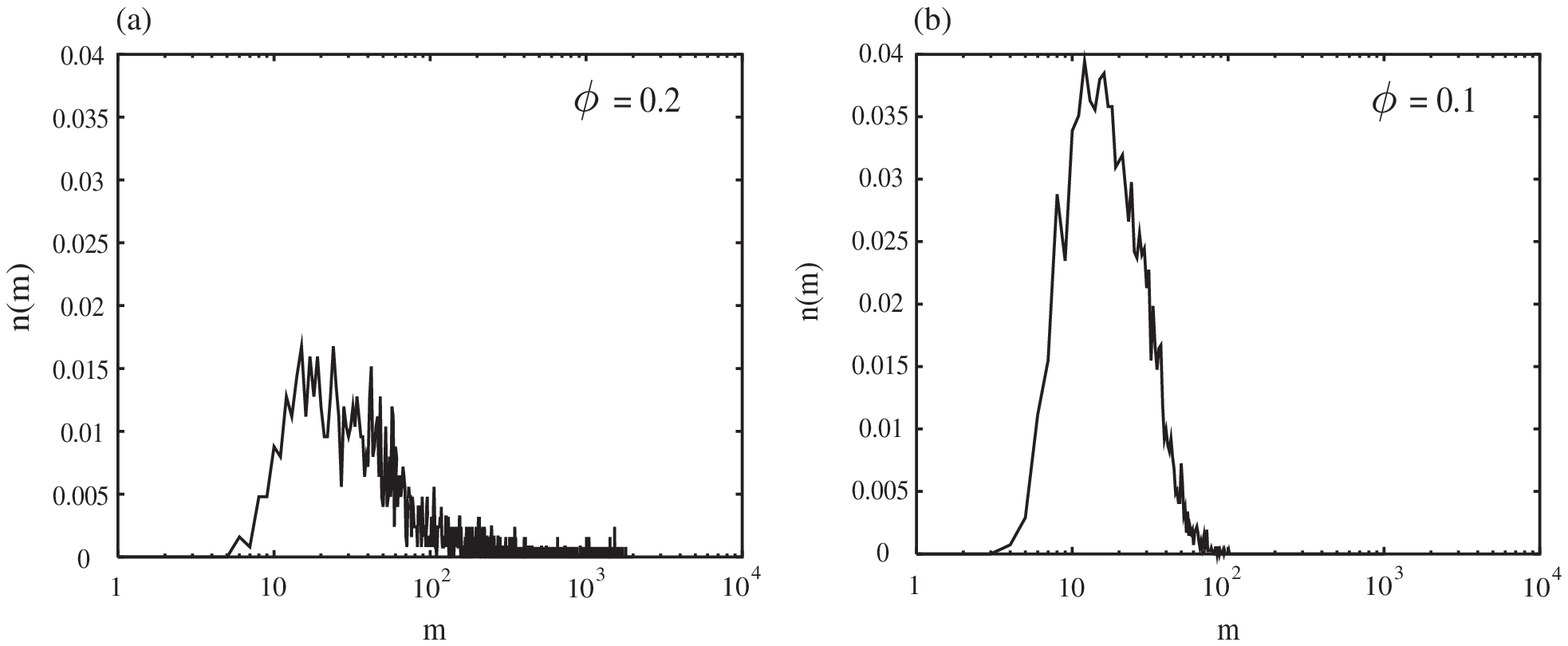}
\caption{Cluster size distribution at $t = 1000$ below the critical
 concentration; (a) $\phi=0.2$ and (b) $\phi=0.1$. 
The vertical axes $n(m)$ denote the number of clusters with mass $m$
 over the total number of clusters.
The broadening distribution appears at $\phi =0.2$. We take data
 from 100 samples.}
\label{fig:c_dis}
\end{figure*}

In order to investigate the structural changes after the gelation
quantitatively, we compare
pore size distributions in the gel state at two different times
$t=t_{\rm g}$ and $t \gg t_{\rm g}$. 
The pore size is defined as the number of sites in the lattice
surrounded by bonding monomers. The precise definition is given as follows.
At first, black dots are put on all sites in the
lattice, and then the dots covered by monomers are changed to
white. Furthermore we define that the bond width is
one unit length of the lattice, and the dots covered by the bonds are also 
changed to white
ones. As a result, we obtain clusters of black dots such as
Fig.\ref{fig:pore_exam}. We define a cluster of black dots as a pore,
and the number
of black dots in the cluster as the pore size $s_{\rm p}$. 

Figure \ref{fig:pore} shows the pore size distributions for 
$\phi = 0.3\ ( > \phi_{\rm g})$ in two system sizes $L =200$ and $400$ 
at two different times; (a) gelation time $t = 10\ ( = t_{\rm g})$ and (b)
$t=1000$.
We can neglect the part in which the data become flat in large 
$s_{\rm p}$ because the flat part depends on system size $L$.
At $t=t_{\rm g}$ the distribution shows two kinds of power law decay
with different exponents. For small $s_{\rm p}$, the distribution
behaves as $P(s_{\rm p}) \sim s_{\rm p}^{-\tau}$ with
$\tau = 1.444 \pm 0.027$, and for large $s_{\rm p}$
 $\tau = 1.982 \pm 0.030$.
The former value of $\tau$ depends on the monomer concentration $\phi$,
but the latter $\tau$ does not.
Since the latter value is near
the Fisher exponent of percolations in two dimensions, 
$\tau_{\rm F} = 187/91 \sim 2.055$ \cite{Stauffer}, our model belongs to
the same universal class in large scales. The distribution is scale
invariant and no inhomogeneity is observed.
On the other hand, the
distribution at $t=1000$ shows a hierarchical structure separated at
$s_{\rm p} \sim s_{\rm pc}$. See Fig.\ref{fig:pore}(b). 
For $s_{\rm p} < s_{\rm pc}$, the distribution shows an exponential
decay, and for $s_{\rm p} > s_{\rm pc}$, it can be fitted by the dashed
line given by 
$P(s_{\rm p}) \sim s_{\rm p}^{-\theta} \exp{(- s_{\rm p}/s_{\rm p}^{\ast})}$ 
with $\theta = 0.581 \pm 0.016$ and 
$\ s_{\rm p}^{\ast} = 427.94 \pm 37.08$. The former comes from small pores in
monomer dense regions in
Fig.\ref{fig:snapshot}(b) at $t=1000$, and the latter from voids in the gel
networks. The scale $s_{\rm p}^{\ast}$ indicates the characteristic size of
the voids. 
The inhomogeneous structure of the gel networks are characterised by two
length scales $\xi_{\rm pc}$ and $\xi_{\rm p}^{\ast}$ corresponding to
$s_{\rm pc}$ and $s_{\rm p}^{\ast}$, respectively
(Fig.\ref{fig:pore_xi}). In comparison with
the distribution at $t=t_{\rm g}$, the distribution at $t \gg t_{\rm g}$ 
contrasts the monomer dense regions with the monomer sparse ones.
This phenomena is similar to spinodal decomposition.

We also confirm the existence of $s_{\rm pc}$ and $s_{\rm p}^{\ast}$
in the thermodynamic limit, $L \rightarrow \infty$.
 In Fig.\ref{fig:pore_size_eff}, we plot
their dependence on the system size
$L$ ($\phi = 0.3,\ t = 1000$). It shows that
\begin{equation}
\begin{matrix}
s_{\rm pc} \simeq 11.678 &  & {\rm for}\ L \ge 150, \\
s_{\rm p}^{\ast} \simeq 445.678 &  & {\rm for}\ L \ge 280.
\end{matrix}
\label{eqn:characteristic_sizes}
\end{equation}
Therefore, these two scales do not disappear in the thermodynamic limit.

\subsection{cluster size distribution near the critical concentration}
\label{subsec:cluster_size}
Now we discuss properties near the critical concentration.  
Figure \ref{fig:c_dis}(a) shows the cluster-size distribution near
the critical concentration ($\phi=0.2$) at $t=1000$.
For comparison, we also plot the distribution at
$\phi=0.1$ in Fig.\ref{fig:c_dis}(b).
The data have been taken from 100 samples.  
We can see that when the concentration $\phi$ is low ($\phi=0.1$),
 the distribution is not broad, but near the critical concentration
($\phi=0.2$), it becomes broader and has a tail toward a
large value of $m$. 
This broadening distribution can be understood as a result of overlapping
of clusters near the critical concentration. 
Namely if the overlapping occurs near the critical concentration, the
distribution of clusters are affected by
processes similar to percolations which interpolate the overlapping clusters.
In general, the percolation processes generally gives a distribution with a
tail toward a large value of $m$
\cite{Stauffer}. 
This broadening behaviour can be interpreted as slow dynamics near the
gel critical point.

\section{Summary and discussions}\label{sec:conclusion}

We examine a chemical gelation model
including aggregations of clusters and bond flexibility in two dimensions. 
The model shows a critical concentration 
$\phi_{\rm g}$ below which no sol-gel transition takes place.
Above the critical concentration, aggregations before gelation show a
fractal structure,
but after the sol-gel transition, a
novel inhomogeneous gel fibre network structure emerges due to bond
fluctuations. From the pore size distributions which characterize the
inhomogeneous structure, we find that the structure can be divided into two
hierarchical structures; monomer-rich region $(r \sim \xi_{\rm pc})$, and 
monomer-poor region $(r \sim \xi_{\rm p}^{\ast})$.
This inhomogeneous structure is important to figure out properties of
the cross-linked gels. 
Experimentally, the inhomogeneous structure can be detected as a speckle
pattern in the light scattering experiment \cite{Tanaka, Shiba3, Zhao} and
a cooperative diffusion of
the hierarchical structure can be observed as the so-called gel
mode \cite{Tanaka2}. 
As is suggested by recent experiments \cite{Zhao}, if the speckle
pattern is caused by large voids in gel networks, the scale of speckle
inhomogeneity is given by $\xi_{\rm p}^{\ast}$. 
At the same time, to observe the speckle pattern in the light scattering
experiment, the speckle inhomogeneity should be larger than the laser
wavelength $\lambda$ \cite{Speckle}, so 
$\xi_{\rm p}^{\ast}$ should be larger than a typical laser wavelength
630 nm. From (\ref{eqn:characteristic_sizes}), this gives a lower bound
of the size of the coarse-grained monomer, which is estimated to be
about 50 nm.
In addition, our simulation predicts that no
speckle pattern is observed just after the gelation and bond fluctuation
is essential to the speckle pattern.
We also find that our model realizes the slow dynamics
near the critical concentration $\phi\simeq \phi_{\rm g}$.

Even though our model shows fractal aggregates similar to the DLCA
model, its gel is very different from that
of the DLCA model.
While our model shows no sol-gel transition below the critical concentration, 
the DLCA model shows a sol-gel transition for any non zero
initial concentrations \cite{Gimel}.
Moreover, our model predicts an inhomogeneous gel network structure, 
but the DLCA model predicts a fractal structure even after the gelation.
Although the gel phase of cross-linked polymers is characterized
by the existence of the so-called gel mode \cite{Tanaka2}, 
the DLCA model can not explain such a specific scale because of its
scale invariant nature due to the fractal structure.
These differences result from the lack of bond fluctuation in the DLCA model.
The DLCA model describes colloidal gels rather than cross-linked
polymer gels.
For example, the growth by the DLCA in three dimensions is predicted to
result in a
power-law increase in cluster radius with time, 
$\langle R\rangle \sim t^{1/d_{\rm f}}$, 
where $\langle R\rangle$ is the average cluster radius
and $d_{\rm f}$ is the fractal dimension.  
This scaling behaviour is observed excellently by recent
experiments of colloidal gels in the International Space Station
\cite{Manley}. 

In our model the dynamics of monomers connected by bonds are Brownian motions
with bond restrictions, and the Brownian force acts on all monomers
equivalently regardless of the sizes of clusters they belong to.
On the other hand, if one assume that less Brownian force acts on
monomers in a larger cluster\cite{Jullien, Jullien2},
the inhomogeneous structure cannot be obtained, since
intra-cluster bond formation cannot proceed in large clusters due to
the low mobility of the constituent monomers. 
Enough Brownian force acting on monomers in large clusters are needed to
form the inhomogeneous structures.

Finally, we would like to point out that the structure of aggregations
depends considerably on the functionality of monomers: The system with
$f=2$ cannot show a sol-gel
transition. Furthermore, in the case of $f=3$, the functionality of
monomers becomes easily saturated during gelation processes, thus the
intra-cluster bond formation important
to the inhomogeneous structure can not proceed enough. Therefore, the
minimal functionality showing the inhomogeneous structure is $f=4$,
which is the same as that of a typical cross-linker $N,N$-methylenebis.

\begin{acknowledgments}
We thank M. Shibayama and T. Karino for helpful discussions. The
 computation in this work has been done using the facilities of the
 Supercomputer Center, Institute for Solid State Physics, University of Tokyo.
\end{acknowledgments}
\bibliography{v6}
\end{document}